\documentclass[12pt,a4paper,twoside]{article}
\usepackage{epsfig,natbib,multirow}
\newcommand{\sekshun}[1]                
	{                               
	\section{#1}                    
	\markboth{Eliche-Moral et al.: Satellite Infall and the
	Growth of Bulges of Spirals }{Eliche-Moral et al.: Satellite Infall and the
	Growth of Bulges of Spirals} 
	}

\newlength{\myVSpace}
\date{}
\author{M.C. Eliche-Moral, M. Balcells, J.A.L. Aguerri, \& A.C. 
Gonz\'alez-Garc\'{\i}a}

\begin{document}

 \title{Satellite Infall and the Growth of Bulges of Spiral Galaxies} 
\maketitle 
\begin{center}
{\textwidth 4 cm
\flushleft  Proceeding of the Joint
European and National Astronomical Meeting, "The Many Scales in the Universe",
held in Granada, Spain, September 13-17, 2004.\/}
\end{center}
\begin{center}

{\textwidth 8 cm
\flushleft \it Instituto de Astrof\'{\i}sica de Canarias, C/ V\'\i a
L\'actea, E-38200 La Laguna, Tenerife, Spain. Contact: mcem@iac.es, 
balcells@iac.es, jalfonso@iac.es, cglez@iac.es\/}
\end{center}
\begin{abstract}
For bulges of spiral galaxies, the concentration, or Sersic index, increases 
with bulge luminosity and bulge-to-disk ratio $B/D$ 
\citep{Andredakis95}.  Does this trend trace the growth 
of bulges via satellite accretion?  And, is satellite infall consistent with 
this trend?  \citet{Aguerri01} (ABP01, hereandafter) investigated 
this question with N-body simulations of the accretion of dense, spheroidal 
satellites.  Here, we expand on that work by running N-body simulations of 
the accretion of satellites that have realistic densities.  Satellites are 
modeled as disk-bulge structures with their own dark-matter halo.  
A realistic density scaling with the primary galaxy is ensured by using 
the Tully-Fisher relation. Our merger models show that most satellites disrupt 
before reaching the center. However, a bulge-disk decomposition of the surface 
density profile after the 
accretion shows an increase of both the $B/D$ and the Sersic index $n$ of the bulge. 
The increase in the mass and concentration of the inner Sersic component is 
due to inward piling up of disk material due to transient bars during the 
satellite orbital decay.  This research is described in \citet{Eliche05}. 
\end{abstract}

\sekshun{N-body models}
\label{Sec:models}
Both the primary and satellite galaxies comprise an exponential disk, 
a King-model bulge and a dark halo built as an Evans model.  
N-body realizations are built following \citet{Kuijken95}.  
The satellite luminous mass scales with the mass of the initial bulge 
as 1:2, 1:3 and 1:6.  Relative sizes, or densities, are determined by 
applying the Tully-Fisher to the primary and secondary.  We experimented 
with $\alpha _{TF}$ exponents of 3, 3.5 and 4.0, although the exponent did 
not affect the main results.  In total, we had 10 experiments. 
Orbital parameters for the merger experiments, satellite mass ratios and
half-mass radii are given in Table \ref{Tab:orbits}.\\

Models were run using a TREECODE from \citet{Hernquist89}, using 185,000 particles 
in each experiment.  Masses, radii and number of particles are given in Table 
\ref{Tab:models}. In order to get structural parametes of the remnants, we 
performed S\'ersic+exponential fits to the face-on 
azymutally-averaged radial surface 
brightness profiles using a code described in \citet{Graham01}. The S\'{e}rsic law 
was used for fitting the bulge 
\citep{Sersic68,Graham01,Mollenhoff01,MacArthur02}:

\begin{equation}
I(r)=I_{e}\cdot exp\,\{ b_{n}\cdot [ ( r/r_{e}) ^{1/n}-1] \}  \label{Eq:Sersic}
\end{equation}

where $r_e$ is the half-light radius, $I_{e}$ is the surface brightness
at $r_e$ and $n$ is the S\'{e}rsic index. The factor $b_{n}$ is
a function of the concentration parameter $n$. An approximation that gives
good results in the range $n<$10 is b$_{n}$=1.9992$\cdot n$-0.3271 
\citep[see][]{Capaccioli87,Graham01}. 
Disk contribution can be fitted with the
exponential law: 
\begin{equation}
I(r)=I_{0}\cdot exp\left( -r/h_D\right)  \label{Eq:disk}
\end{equation}

where $h_D$ is the disk scale length and $I_0$ is its central surface
brightness.\\

\begin{figure}[h!]
\begin{center}
\includegraphics[width=0.9\textwidth,angle=0]{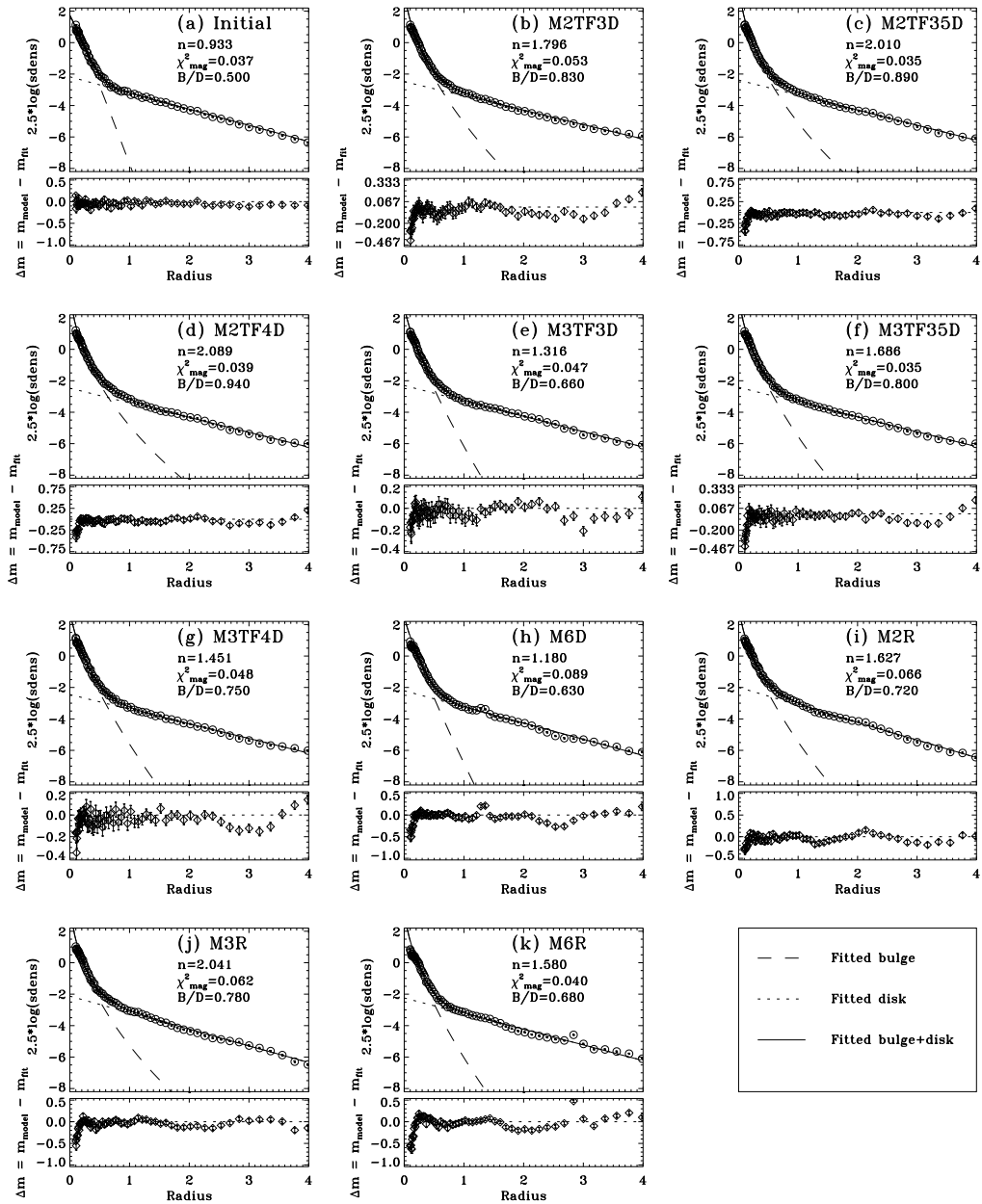}
\end{center}
\caption{Radial surface density profiles of luminous matter and
S\'ersic+exponential simultaneous fits. {\it Upper panels of each frame\/}: 
Surface density distributions and performed fits. Panels: ({\it a\/}) Initial
model. ({\it b-h\/}) Prograde models after the merger is complete. 
({\it i-k\/}) Retrograde
models after the merger is complete. {\it Open circles\/}: model
measurements. {\it Dashed line\/}: S\'ersic $r^{1/n}$ fitted component. 
{\it Dotted lines\/}: Exponential fitted component. 
{\it Solid lines\/}: Sum of the two fitted components. 
{\it Lower panels of each frame\/}: Residuals in magnitudes. Error bars 
of model measurements in 
magnitudes are plotted upon the residual points ({\it diamonds\/}).}
\label{Fig:sbr}
\end{figure}

The final face-on,
azymutally-averaged radial surface density profiles of the luminous matter
for all the models are shown in Figures \ref{Fig:sbr}b-k. Figures \ref{Fig:sbr}a is
the surface density profile of the luminous matter for the primary galaxy.
Dotted and dashed lines are the simultaneous two component fits
perfomed to the total luminous surface density (exponential plus
S\'{e}rsic-law). Residuals in magnitudes of the
fits appear down its corresponding surface brightness profile in 
Figure \ref{Fig:sbr}. As can be seen, they are less than 0.15 mags in all 
the cases, a 
quite reasonable result compared to typical observational errors. Final
fitted parameters and bulge-to-disk mass ratios derived from the fits
are tabulated in Table \ref{Tab:fits}.

\begin{table}[bh!]
\caption{Initial parameters of the primary galaxy and the satellites.}\label{Tab:models}
\begin{center}\tiny
\begin{tabular}{cccccccccccccccc}
\hline \hline\vspace{0.2cm}
\multirow{2}*{} & &\multicolumn{3}{c}{\multirow{2}*{Primary
Galaxy}}  & &\multicolumn{3}{c}{\multirow{2}*{Satellites}} & & 
\multicolumn{6}{c}{\multirow{2}*{Primary Galaxy
Characteristics}}\\[0.1cm]
\cline{3-5}\cline{7-9}\cline{11-16}\\
 NP & & Disk1 & Bulge1 & Halo1 & &  Disk2 & Bulge2 & Halo2 & & $M_{Bulge}$& $M_{Disk}$ 
  & $M_{Dark}$ & $r_{B}$ & $h_{D}$ & $z_{D}$ \\
 (1)& & (2)  & (3)    & (4)  & & (5)   & (6)    & (7) &  & (8)  & (9)  & (10) 
&(11) &(12) &(13)\\[0.2cm]\hline\\[-0.2cm]
 185K & &40K & 10K & 90K & & 10K & 5K & 30K & &0.42 & 0.82 & 5.20 & 0.195 &
 1.0&0.1\\[0.2cm]\hline
\end{tabular}
\end{center}
{\it Column description\/}: (1) Total particle No.\ (2) No.\ of primary disk particles. 
(3) No.\ of primary bulge particles. (4) No.\ of primary halo particles. 
(5) No.\ of satellite disk particles. (6) No.\ of satellite bulge particles. 
(7) No.\ of satellite halo particles. (8) Primary bulge mass. (9) Primary 
disk mass. (10) Primary halo mass. (11) Primary bulge half-mass radius. (12) Disk
truncation radius. (13) Disk scale height.
\end{table}

\vspace{0.5cm}
\begin{table}[bh!]
\caption{Orbital and scaling parameters for the merger experiments.}\label{Tab:orbits}
\begin{center}\scriptsize
\begin{tabular}{ccccccccc}

\hline\hline\\[-0.2cm]
Model & Code & $M_{Sat}(Lum)/M_{G}(Bulge)$ & $M_{Sat}/M_{G}$  &
$\alpha _{TF}$ & $R_{Sat}/R_{G}$ & $V_{R}$ & $V_{\theta}$ & $\theta _{1}$ \\
 (1)&  (2)  & (3)    & (4)  & (5)   & (6)    & (7) &  (8) & (9)\\[-0.2cm]\\\hline\\[-0.2cm]
(1)  & M2TF4D & 1/2 & 0.16 & 4.0  & 0.4  & -0.00142 & 0.24873 & 30 \\
(2)  & M2TF35D& 1/2 & 0.16 & 3.5  & 0.46 & -0.00142 & 0.24873 & 30 \\
(3)  & M2TF3D & 1/2 & 0.16 & 3.0  & 0.54 & -0.00142 & 0.24873 & 30 \\
(4)  & M3TF4D & 1/3 & 0.11 & 4.0  & 0.33 & -0.00131 & 0.24331 & 30 \\
(5)  & M3TF35D& 1/3 & 0.11 & 3.5  & 0.39 & -0.00131 & 0.24331 & 30 \\
(6)  & M3TF3D & 1/3 & 0.11 & 3.0  & 0.48 & -0.00131 & 0.24331 & 30 \\
(7)  & M6D    & 1/6 & 0.05 & 3.5  & 0.28 & -0.00029 & 0.23664 & 30 \\
(8)  & M2R    & 1/2 & 0.16 & 3.5  & 0.46 & -0.00142 & 0.24873 & 150 \\
(9)  & M3R    & 1/3 & 0.11 & 3.5  & 0.39 & -0.00131 & 0.24331 & 150 \\
(10) & M6R    & 1/6 & 0.05 & 3.5  & 0.28 & -0.00029 & 0.23664 & 150
\\\hline
\end{tabular}\end{center}
{\it Column description\/}: (1) Model number. (2) Model code. 
(3) Initial mass ratio between luminous satellite material and primary bulge
material. (4) Initial mass ratio between satellite and primary galaxy.
(5) Tully-Fisher index for scaling. (6) Initial half-mass radius ratio between
satellite and bulge. (7) and (8) Radial and tangential velocity components of
the relative orbit. (9) Initial angle between the orbital momentum and the
primary disk spin. The other three angles involved in the orbits are fixed: 
$\phi _1$=0$^\circ$, $\theta _2$=25$^\circ$, and $\phi _2$=90$^\circ$.
\end{table}

\vspace{0.5cm}
\begin{table}[b!h]
\caption{Fitted parameters for S\'ersic bulge plus exponential disk decomposition of the
final remnants.}\label{Tab:fits}
\begin{center}\scriptsize
\begin{tabular}{ccccccccccccc}

\hline\hline\\[-0.2cm]
   &   & 
 & &\multicolumn{2}{c}{Disk}  & &
\multicolumn{3}{c}{Bulge}& &\\
\cline{5-6}\cline{8-10}\\[-0.2cm]
Model&Code &$\chi ^2$ (mag) &&$\log(\mu _0)$ & $h_D$  & &
$\log(\mu _e)$ &  $r_e$  & $n$ & &$B/D$\\
 (1)&  (2)  & (3) &   & (4)  & (5)   & &(6)    & (7) &  (8)  & 
 &(9)\\[-0.2cm]\\\hline\\[-0.2cm]
Initial&...&0.037&&-0.87$\pm$0.01 & 1.05$\pm$0.02 && 0.01$\pm$0.04 &
0.199$\pm$0.02 & 0.92$\pm$0.21&& 0.50\\
(1)& M2TF3D&0.053&&-0.99$\pm$0.05 & 1.19$\pm$0.03 && 0.14$\pm$0.02 &
  0.187$\pm$0.03 & 1.80$\pm$0.16
  && 0.83\\
 (2)& M2TF35D&0.035&&-0.95$\pm$0.02 & 1.13$\pm$0.02 && 0.23$\pm$0.06 & 0.169$\pm$0.01 & 2.01$\pm$0.23
 && 0.89\\
  (3)& M2TF4D&0.039&&-0.99$\pm$0.02 & 1.16$\pm$0.05 && 0.18$\pm$0.01 &
  0.179$\pm$0.03 & 2.09$\pm$0.15
  && 0.94\\
  (4)& M3TF3D&0.047&&-0.92$\pm$0.03 & 1.11$\pm$0.04 && 0.13$\pm$0.02 &
  0.185$\pm$0.03 & 1.32$\pm$0.09
  && 0.66\\
(5)&  M3TF35D&0.035&&-0.97$\pm$0.01 & 1.15$\pm$0.06 && 0.21$\pm$0.03 &
 0.172$\pm$0.02 & 1.69$\pm$0.11
 && 0.80\\
  (6)& M3TF4D&0.048&&-0.98$\pm$0.02 & 1.17$\pm$0.05 && 0.11$\pm$0.01 & 0.193$\pm$0.05 & 1.45$\pm$0.11
  && 0.75\\
     (7)& M6D&0.089&&-0.88$\pm$0.07 & 1.06$\pm$0.03 && 0.14$\pm$0.02 & 0.181$\pm$0.04 & 1.18$\pm$0.13
     && 0.63\\
    (8)&  M2R&0.066&&-0.80$\pm$0.04 & 0.97$\pm$0.02 && 0.13$\pm$0.04 &
     0.184$\pm$0.03 & 1.63$\pm$0.17
     && 0.72\\
    (9)&  M3R&0.062&&-0.85$\pm$0.04 & 1.03$\pm$0.06 && 0.23$\pm$0.05 & 0.162$\pm$0.04 & 2.04$\pm$0.13
     && 0.78\\
    (10)&  M6R&0.040&&-0.90$\pm$0.01 & 1.09$\pm$0.07 && 0.22$\pm$0.02 & 0.163$\pm$0.04 & 1.58$\pm$0.12
     && 0.68\\\hline
\end{tabular}\end{center}
{\it Column description\/}: (1) Model number. (2) Model code. 
(3) $\chi ^{2}$ of the fit. (4) Disk central intensity. (5) Disk scale length.
(6) Bulge effective surface density. (7) Bulge effective radius. (8) Bulge profile
S\'ersic index. (9) Bulge-to-disk mass ratio derived from the S\'ersic+exponential
fit.

\end{table}

\sekshun{Growth of bulges}
\label{Sec:growthofbulges}

\begin{figure}[h!]
\begin{center}
\includegraphics[width=0.95\textwidth,angle=0]{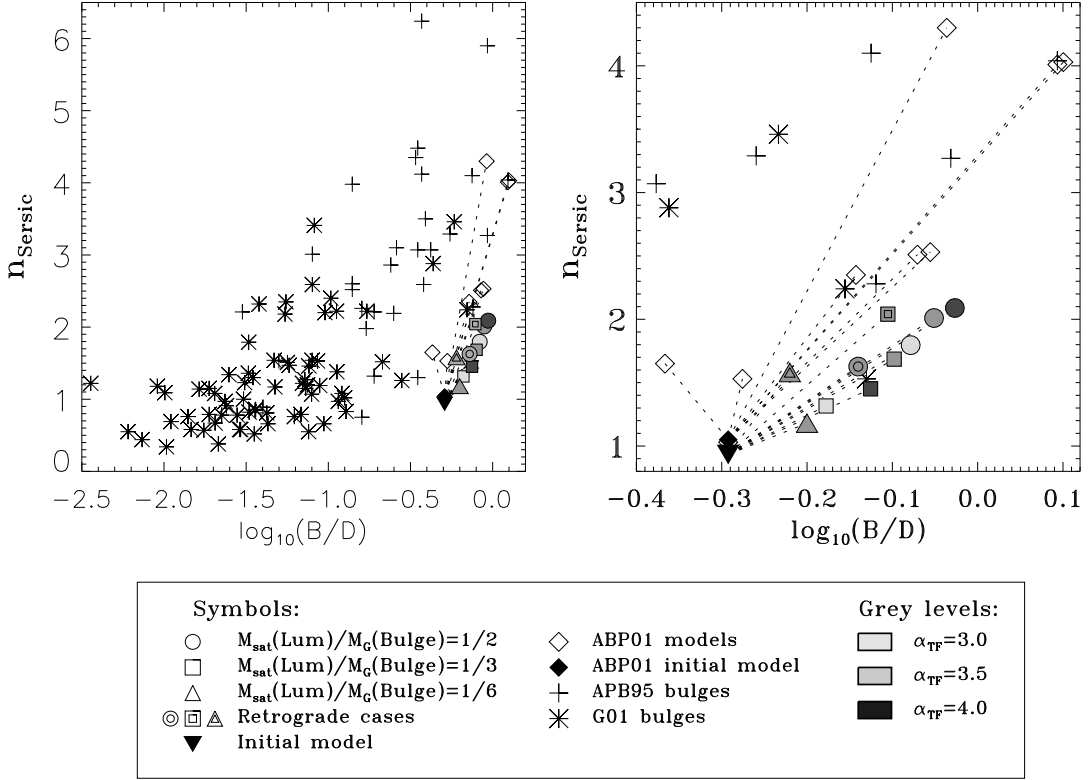}
\end{center}
\caption{Growth vectors in the $n$-log\,($B/D$) plane for the present
low-density models. Right panel shows a zoom of the region from the 
left panel where our models are. Each arrow starts at the location of the original model
and ends at the $n$ and $B/D$ derived from the two-component fit to the
surface density profile after the merger. Growth vectors of ABP01
high-density models are plotted for comparison. Distributions of $n$ 
vs.\,log($B/D$) for bulges of several studies are drawn too. Grey colour level of
filled points indicates the $\alpha _{TF}$ exponent used for each
simulation: the palest grey for $\alpha _{TF}$=3, the darkest grey for 
$\alpha _{TF}$=4 and the intermediate grey level for $\alpha _{TF}$=3.5. 
{\it Inverse filled triangule\/}: Initial model. {\it Colored
circles\/}: Mass ratio 1:2. {\it Colored squares\/}: Mass ratio 1:3. 
{\it Colored triangles\/}: Mass ratio 1:6. {\it Double centered symbol\/}: 
Retrograde models. {\it Crosses\/}: Observed
bulges from APB95. {\it Asterisks\/}: Observed bulges from 
\citet{deJong96b}, re-analised by G01. {\it Filled diamond point\/}
: Initial model of ABP01. {\it Empty diamonds}: ABP01's remnants after the
high-density satellite accretions.}
\label{Fig:growth}
\end{figure}

Figure \,\ref{Fig:growth} show growth vectors in the plane $n$ vs.\,log\,($B/D$), where the $B/D$ ratios are derived from the
S\'{e}rsic+exponential fits. Plotting points indicate the characteristics
of our run models, according to the legend in the Figure. Real bulges from the
samples of \citet{deJong96b} (re-analysed by G01) and APB95 are drawn too,
together with the growth vectors of ABP01 high-density models (diamond points)
for comparison. Growth vectors in the plane $n$ vs.\, $B/D$ show a similar 
dependence to the one found by ABP01: $\mu(r)$ evolves quickly from an initial 
exponential bulge $n\sim$1 to earlier types bulges in all the cases, reaching $n$=2.1, 
and proportionally to the satellite mass.
Then, not only high-density, but also low-density satellite accretion onto 
disk-bulge-halo galaxies causes the bulge surface brightness profile  to 
evolve toward higher-$n$ Sersic profiles, following similar increasing trends 
and values for the Sersic index $n$ with $B/D$ ratios just as in the 
observations.  The low-density experiments fill the region in the plane $n$
vs.\, $B/D$ that ABP01 high-density experiments left empty.\\

\sekshun{Why does $n$ increase}
\label{Sec:whynincreases}

\begin{figure}[!]
\begin{center}
\includegraphics[width=0.85\textwidth,angle=0]{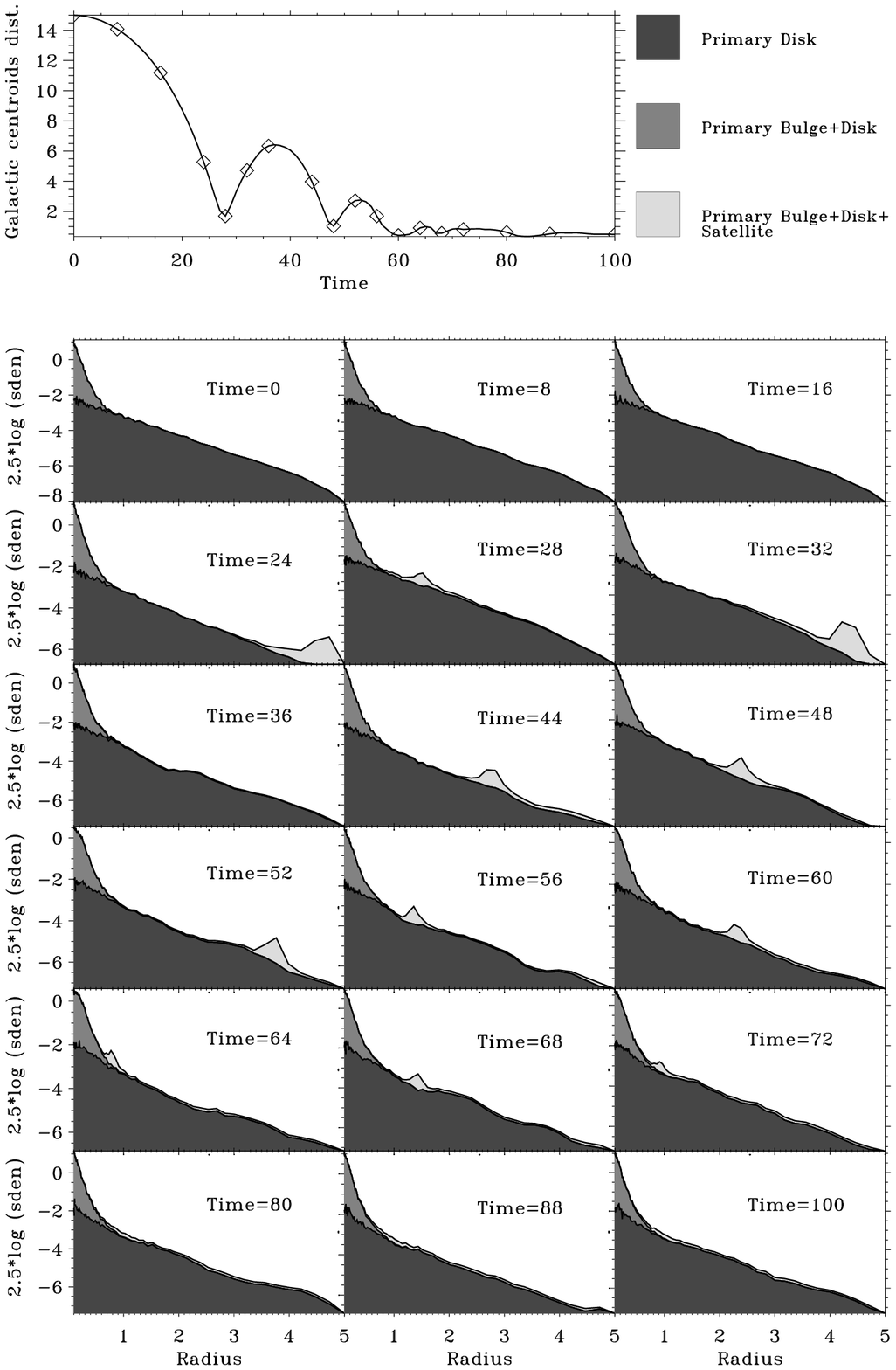}
\end{center}
\caption{Evolution of the surface density profiles of the various luminous
components. Contributions of each component to the total profile are drawn
with different grey scales. Time is shown at upper right-corner of each
frame. The first top panel shows the distance between 
centroids of the primary galaxy and the satellite as a function of time. 
{\it Darkest grey-filled region\/}: Contribution of the primary disk
particles to the surface density profile at each time. {\it Intermediate
grey-filled region\/}: Contribution of the primary bulge particles summed up 
to that of the disk primary material. {\it Palest grey-filled region\/}:
Contribution of the satellite luminous material summed up to the luminous
material of the primary galaxy.}
\label{Fig:sbrt}
\end{figure}

The evolution of the bulge profiles in the high-density experiments 
of ABP01 was driven by the puffing up of the bulge material and the 
deposition of the satellite's high density cusp in the remnant center. 
The dynamical mechanism for the increase of $n$ in low-density models is 
different. In Figure \,\ref{Fig:sbrt} we show the time evolution of the primary 
disk and bulge particles 
to the surface brightness profile of the merger in model M3TF3D. 
The contribution of the satellite 
particles is plotted too. The satellite disrupts completely before reaching 
the galactic center, in such a way that this deposition over the remaining 
disk and the injection of disk material to the center are the responsible 
of the steepening of the profile.\\

\begin{figure}[h!]
\begin{center}
\includegraphics[width=\textwidth,angle=0]{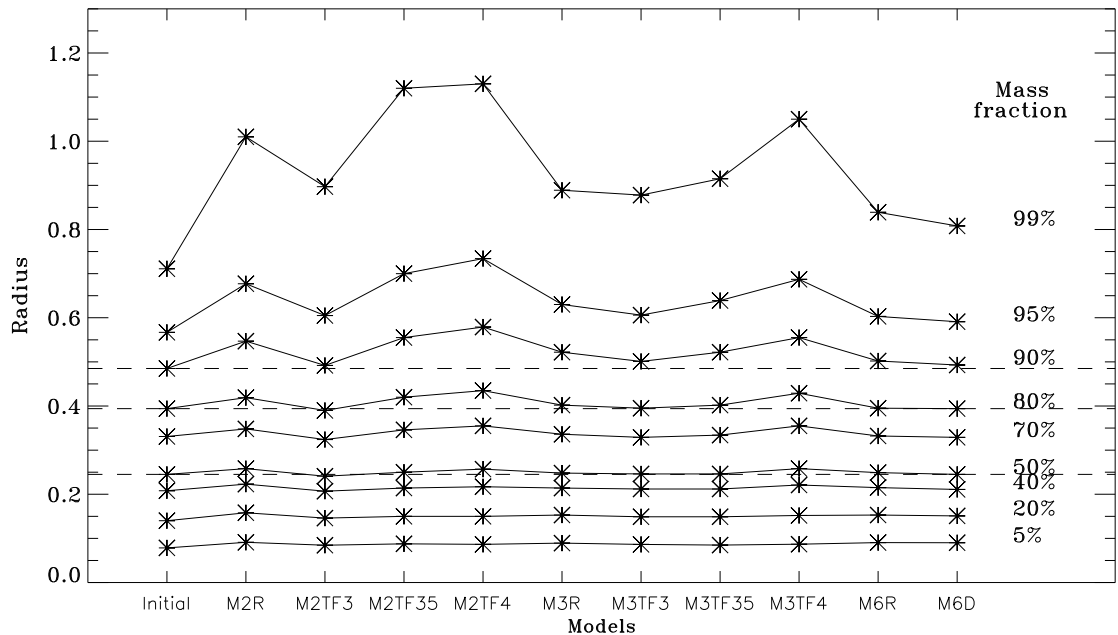}
\end{center}
\caption{Radii enclosing a given \% of the mass for the distribution 
of particles initially belonging to the bulge. The abcissa are the model 
codes from Table \ref{Tab:orbits}. The horizontal dotted lines are the 
radii initially enclosing a given \% of the bulge mass, which appears next 
to the line, on the right of the Figure.}
\label{Fig:bulge}
\end{figure}

Figure \,\ref{Fig:bulge} shows the radii enclosing a given percentage of
the mass for the particles initially belonging to the primary bulge for all
the models. Our bulges basically expand their outer layers, while the 90\% 
of their masses remains undisturbed. Then, bulge material
puffing up can not be the responsible of the increasing of $n$ in our
low-density satellite accretions, contrary to what happened in ABP01's
high-density experiments.\\

\begin{figure}[h!]
\begin{center}
\includegraphics[width=\textwidth,angle=0]{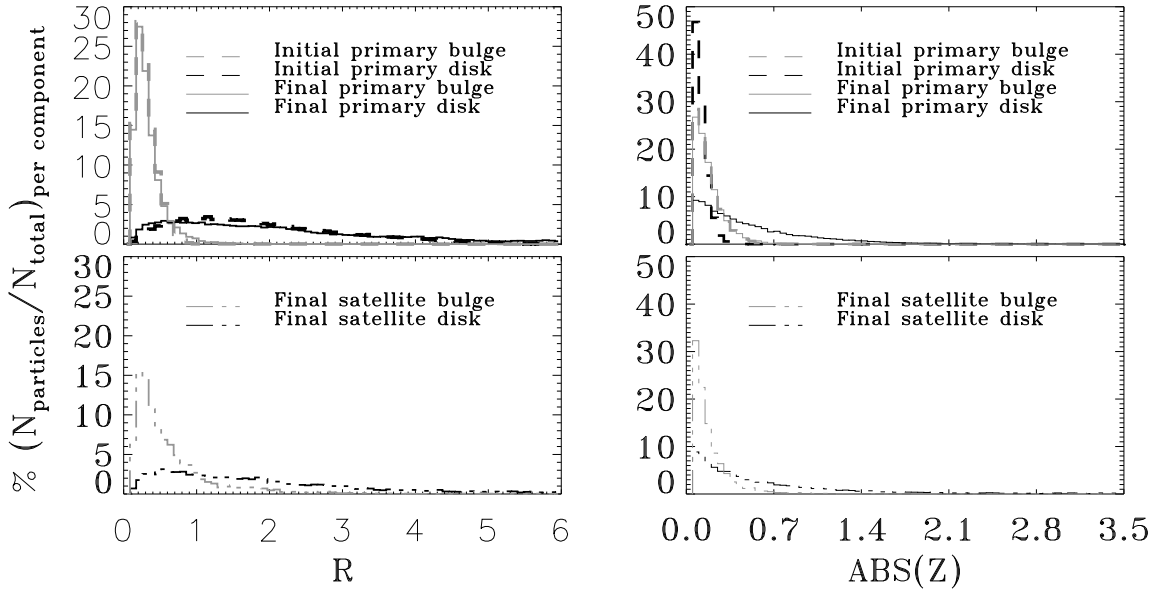}
\end{center}
\caption{Distribution of material of each component in the final remnant of model 
M2TF4D. {\it Left panels\/}: Radial distribution of particles from each component in 
\% respect to the total number which initially belonged to the same 
component. {\it Right panels\/}: The same as left panels, but vertically. 
{\it Dashed lines\/}: Initial distributions of disk and bulge 
particles from the primary galaxy. 
{\it Solid lines\/}: Final distributions of disk and bulge 
particles from the primary galaxy.
{\it Dashed-dotted lines\/}: Final distributions of bulge and 
disk particles from the satellite in the remnant.
Bulge and disk particles are colored by light and dark grey respectively 
in the three cases.}
\label{Fig:contributionrz}
\end{figure}

The distribution of material of each component in the final remnant of model 
M2TF4D is shown as a function of radius and height respect to the
galactic plane in Figure \,\ref{Fig:contributionrz}. In the left panels, 
we have plotted the percentage respect to the
total number of particles which initially belonged to each component at each
galactic radius, $r$. Right panels represent the same as a function of the 
absolute value of the $z$-component. We have separated primary distributions 
(upper panels) from satellite contributions (lower panels) for clarity. 
Dashed lines indicate how the initial bulge and disk 
material were distributed, while solid lines correspond to the final 
acquired distribution by each of them in the upper panels. Dashed-dotted 
lines in the lower ones show final contributions of the satellite's bulge and 
disk material. Bulge and 
disk particles are differenciated by the grey scale (light and dark grey 
respectively) in both satellite and primary galaxy distributions. 
Looking at the upper panels, it is obvious that the distribution of particles 
associated to the primary bulge experiments little changes radially and 
vertically. Primary disk matter experiments an inward flow, as it can be noticed
from Fig.\,\ref{Fig:sbrt}. Their outer 
layers generate tails and
expand due to tidal forces in the direct orbits, and are inhibited in the 
retrograde cases \citep{Mihos96}.
Satellite disk
material rebuilds an exponential disk, following similar trends along the
same range of R and $\left| z\right| $ than the primary disk material.
However, satellite's bulge particles are confined to the remnant's inner
region and to small values of $\left| z\right| $; i.e., satellite bulge material 
contributes to the thin disk. This effect is due to the
fact that dynamical friction circularizes the orbit of the satellite prior
to disruption: stars at the core of the satellite are more resilient to
disruption and therefore end up on more circular orbits than those stripped
earlier during the satellite decay process. As in \citet{Abadi03a,Abadi03b}
simulations, most stars in the satellite are dispersed into a torus-like
structure, whose radius is that at which final disruption takes place. 
This redistribution of material 
produces a population mixture which could be the responsible of the similar 
colors observed between bulges and disks \citep{Peletier96}.

\sekshun{Scaling relations of disks and bulges}
\label{Sec:scaling} 

\citet{deJong96a} and \citet{Courteau96} observed that the 
ratio $r_e/h_D$ between the bulge effective radius and the disk scale 
length is independent of Hubble type, and claimed that the spiral Hubble 
sequence is "scale-free". \citet{Balcells04b} (BGP04, hereandafter) confirm the 
independence of most disk-bulge structural parameters with Hubble type, but 
found that such photometric parameters of bulges and disks strongly correlate 
with bulge luminosity and with Sersic index $n$.  They conclude that galaxies 
themselves are not scale-free, the luminosity of the bulge being the critical 
scale. \\

\begin{figure}[!]
\begin{center}
\includegraphics[width=0.75\textwidth,angle=0]{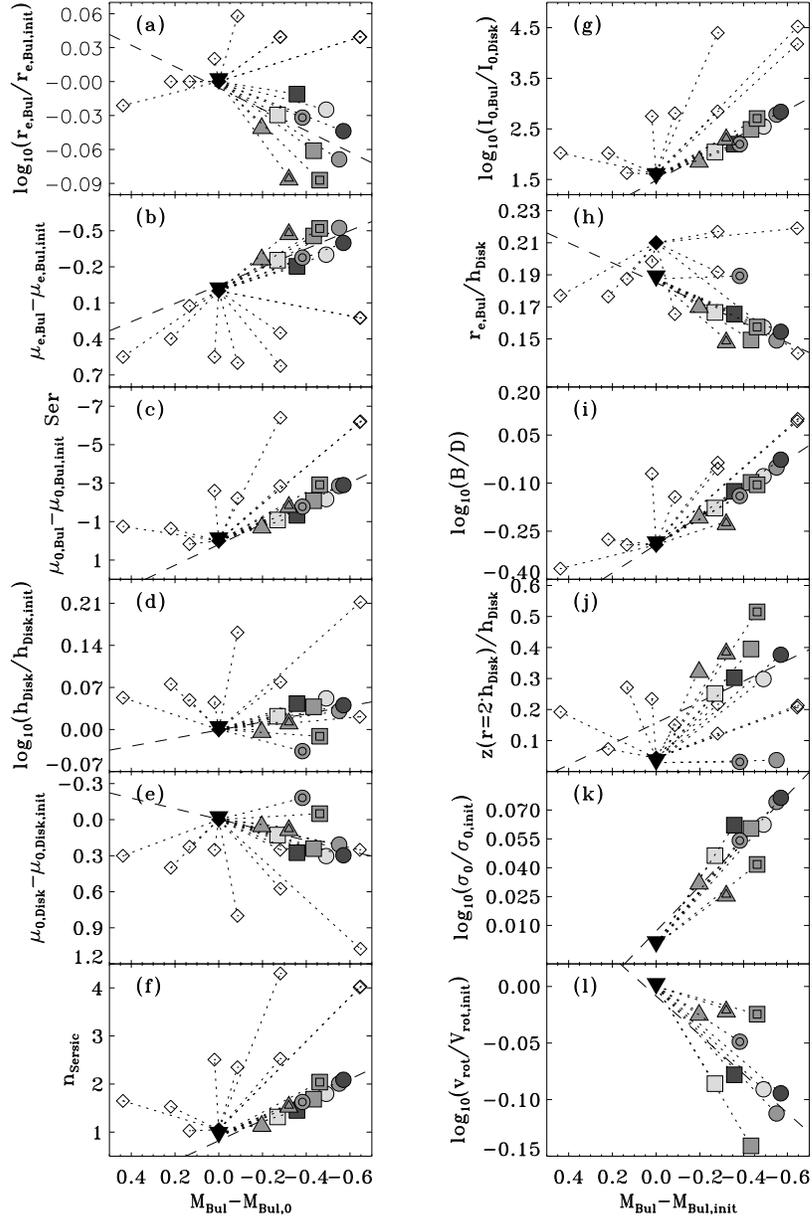}
\end{center}
\caption{The dependence of bulge and disk parameters, and $B/D$ ratios, on 
the bulge "magnitude", defined as 
$M_{Bul}\equiv -2.5\cdot \log\,(Mass_{Bul})$. All parameters are derived 
from Table \ref{Tab:fits}. Legend is the same as in Figure \,\ref{Fig:growth}. 
{\it Dashed lines\/}: Orthogonal regressions to the model points. Obtained 
slopes are compared to those from observations in Table \ref{Tab:slopes}.
({\it a\/}) Effective radius of the (S\'ersic) bulge component. 
({\it b\/}) Effective surface brightness of the S\'ersic component. 
({\it c\/}) Extrapolated central surface brightness of the S\'ersic 
component. ({\it d\/}) Disk major-axis scale length. 
({\it e\/}) Face-on extrapolated disk central surface brightness.
({\it f\/}) S\'ersic index $n$. 
({\it g\/}) Bulge-to-disk central brightness ratio 
log\,($I_{0,Bul}/I_{0,Disk}$). 
({\it h\/}) Ratio $r_e/h_D$ between the bulge effective radius and the 
major-axis disk scale length. ({\it i\/}) Bulge-to-disk luminosity ratio 
$B/D$. ({\it j\/}) Ratio $z_D(r=2\cdot h_D)/h_D$ between the disk 
height scale at $r$=2$\cdot h_D$ and the major-axis disk scale. 
({\it k\/}) Central velocity dispersion. ({\it l\/}) Maximum rotational 
velocity.}
\label{Fig:parameters}
\end{figure}

In Figure \,\ref{Fig:parameters}, the dependences of the bulge and disk
photometric parameters on the mass of the final bulge are shown, this last
parameter obtained from the simultaneous S\'ersic-exponential fits performed
to the surface density profiles of the remnants. Supposing that the
mass-to-light ratio $M/L$ is constant and very similar for real bulges 
\citep{Portinari04}, we have defined the bulge magnitude as 
$M_{Bul}\equiv -2.5\cdot \log\,(Mass_{Bulge})$, because the bulge total mass can be directly
related to its luminosity through the constant $M/L$. We have plotted ratios
between the parameters of the final merger remnants and the corresponding to
the initial galaxy vs.\ the increment in the bulge magnitude 
(i.e., final minus initial magnitude). Legend for symbols is the
same as in Figure \,\ref{Fig:growth}. Notice that the strong proportional
relations between photometric parameters and the bulge magnitude in our
low-density models are weaker in the ABP01's dense models (represented by
empty diamonds in the Figure); trends as those exhibited by $\mu _{0,Bul}$, 
$n_{S\acute{e}rsic}$, log\,($I_{0,Bul}/I_{0,Disk}$), log\,($B/D$) and 
log\,($\sigma _0$) (see panels (c), (f), (g), (i) and (k) in the Figure) 
correspond to correlation coefficients greater than 0.96 when a lineal fit 
is performed (you can 
see these values in Col.\,6 from Table \ref{Tab:slopes}). 
This means that low-density satellite accretion
processes do not alter galaxies randomly: they give place to remnants whose
properties are scaled between them, depending on the mass ratios and the
orbit of the encounter. In adddition, all the remnants have brighter bulges
than initially.\\
BGP04 gave mathematical expressions for the stronger relations, as follows: 
\begin{equation}
\log\,(H)=\left( m\pm \Delta m\right) \cdot M_{K,Bulge} + \left( n\pm \Delta
n \right)  \label{Eq:correlation}
\end{equation}

where $H$ represents a given photometric parameter (as $\mu _{0,Bulge}$, 
$\mu_{0,Disk}$, $r_e$, $h_{D}$, $n$...), or the ratio between two of them; 
$M_{K,Bulge}$ is the K-band bulge magnitude; and ($m\pm\Delta m$) and 
($n\pm\Delta n$) are the slope and zero-point obtained from the orthogonal
regression to the log($H$)-$M_{K,Bulge}$ relation. We have performed linear
fits to the relations shown at Figure \,\ref{Fig:parameters}, representing
results by the dashed line in each case. Therefore, if real bulges grow
through satellite accretions, our slopes should be similar to those found by
BGP04, because our models are scalable and hence they can be displaced in
the plane log($H$)-$M_{K,bulge}$. The constants needed for 
transforming from masses to luminosities and from a unit system to another 
do not affect the slopes in the
relations, as they end as part of the additive term of the linear fit. \\

In Table \ref{Tab:slopes}, the slopes for all the strong
correlations observed by BGP04 are compared to those obtained in our
low-density models. It is encouraging that low-density models reproduce the
observed tendencies in disk parameters, central velocity dispersions and 
$B/D$ ratios. The correlations which we fail to reproduce with our experiments
are those involving the bulge effective radius $r_e$ (see $r_e$ in panel 
(a) and $r_e/h_D$ in panel (h) in the Figure). 
$r_e$ becomes smaller
with the accretion, while observed higher luminosity bulges have larger
effective radii 
\citep[][ among others]{Hubble26,Binggeli84,Mollenhoff01}. 
This is due to the fact 
that the primary bulge material remains unaltered, while the bulge region 
receives a large inward piling up of  disk particles that rises the central 
galactic brightness, making the effective radius decrease. 
On the other hand, the slope for the bulge central brightness 
$\mu_{0,Bulge}$ is very different from the observed value also, despite its sign
is positive in both cases (see Table \ref{Tab:slopes}). Moreover, we find a 
correlation between the bulge effective brightness and L$_{K,bulge}$, while
BGP04 do not detect a clear tendency in their sample of intermediate- to
early-type spirals. Of course, these discrepancies remain important
limitations of the current accretion models. Probably, these problems are 
associated to the fact that models without
star formation and gas hydrodynamics implemented give an incomplete vision
of the physical processes involved in the galactic accretions. These 
processes are known to affect central structure of real galaxies
significantly \citep{Mihos96}. Hence, all the relations 
found for the bulge
parameters and ratios related to them must be distrusted. 
On the other hand,
correlations involving large-scale parameters of the galaxy are well
reproduced, probably because they are less affected by star formation.\\

\begin{table}[h!]
\caption{Photometric parameters slopes vs.\, bulge magnitude from our models and
from BGP04 observations.}\label{Tab:slopes}
\begin{center} \scriptsize
\begin{tabular}{lr@{$\pm$}lr@{.}lclr@{$\pm$}lr@{.}lr@{.}l}

\hline\hline\\[-0.2cm]
\multicolumn{5}{c}{Observational}&\multicolumn{7}{c}{Modelled}\\
\cline{1-5}\cline{7-13}\\[-0.2cm]
\multicolumn{1}{c}{Photometric parameter} &  \multicolumn{2}{c}{Obs.
Slope}
 & \multicolumn{2}{c}{$R_S$}  && 
\multicolumn{1}{c}{ Photometric parameter } & 
 \multicolumn{2}{c}{Model Slope}  &
 \multicolumn{2}{c}{$R_S$} 
&\multicolumn{2}{c}{$\chi ^2$} \\
\multicolumn{1}{c}{(1)} &  \multicolumn{2}{c}{(2)}
 & \multicolumn{2}{c}{(3)}  && 
 \multicolumn{1}{c}{(4)} & 
 \multicolumn{2}{c}{ (5)}  &
 \multicolumn{2}{c}{(6)} 
&\multicolumn{2}{c}{(7)}
\\[-0.2cm]\\\hline\\[-0.2cm]
log\,($r _e/Kpc$)             	  & -0.164&0.028& -0&79 &&log\,($r_e/r_{e,inic}$)     	&0.09 &0.04&    0&41&
0&004\\
$\mu _{e,K} /M _{K,\odot}$       & \multicolumn{2}{c}{no correlation} & \multicolumn{2}{c}{...}&&$\mu _e- \mu_{e,inic}$ 	& 0.76 &0.19&	0&67&
0&068\\
$\mu _{K,Ser} (0)/M _{K,\odot}$  &0.92&0.16 &0&63 &&$\mu _{0,B} - \mu _{0,B,inic}$ 	& 5.4  &0.6&	0&98&
0&713\\
log\,($h _D/Kpc$)          	  &-0.112&0.016 & -0&64&&log\,($h_D/h_{D,inic}$)	& -0.07 &0.04&   -0&73&
0&003\\
$\mu _{K,D} (0)/M _{K,\odot}$ 	  & -0.30&0.06& -0&63&&$\mu _{0,D} - \mu _{0,D,inic}$	&-0.44&0.22&   -0&80&
0&094\\
log\,($I_{0,B}/I_{0,D}$)    	  &\multicolumn{2}{c}{not
significative}&\multicolumn{2}{c}{...} &&log\,($I_{0,B}/I_{0,D}$)	&-2.33&0.17 &  -0&99&
0&054\\
log\,($r _e/h _D$)               	  &\multicolumn{2}{c}{no correlation}
&\multicolumn{2}{c}{...} &&log\,($r _e/h _D$)		&0.062&0.008&   0&87&
0&000\\
log\,($B/D$)	             	  &-0.30&0.04 &-0&80 &&log\,($B/D$)			&-0.442&0.018&  -0&99&
0&001\\
$n_{S\acute{e}rsic}$	             	  &\multicolumn{2}{c}{not significative} &
\multicolumn{2}{c}{...}&&$n_{S\acute{e}rsic}$			&-2.13&0.23&
-0&99&   0&107\\
...	           		  & \multicolumn{2}{c}{...}& \multicolumn{2}{c}{...}&&$z_D/h_D$		   		&-0.34&0.09&    -0&82&   0&157\\
log\,($\sigma _0/km\, s^{-1}$)	  &-0.13&0.02 &-0&81 &&log\,($\sigma _0/\sigma_{0,inic}$) &   -0.118&0.019&  -0&96&
0&001\\
... 						&\multicolumn{2}{c}{...}
&\multicolumn{2}{c}{...}& &log\,($V _{rot}/V _{rot,inic}$)  	 &0.17&0.07&    0&67&
0&009\\\hline

\end{tabular}\end{center}
{\it Column description\/}: Columns (1)-(3), fitted slopes of the 
correlations between photometric
parameters and the K-band bulge magnitude, from the observational sample 
of intermediate- to early-type spirals by BGP04. Columns (4)-(7), 
fitted slopes for the equivalent correlations obtained from our models, using 
the corresponding structural parameters.\\
{\it Columns\/}: (1) Observational photometric parameter.
(2) Observational fitted slopes of parameter in Col.\,1 vs.\ $M_{K,Bul}$ from BGP04. 
(3) Observational correlation coefficient for the fit. 
(4) Structural parameters from our low-density models.
(5) Fitted slopes of parameters in Col.\,4 vs.\ $M_{Bul}-M_{Bul,0}$.
(6) Correlation coefficient for the fit. 
(7) $\chi ^2$ of the fit.

\end{table}

\sekshun{Summary}
\label{Sec:summary}

The study of the effects of satellite infall onto galaxies needs to 
consider the dynamical transformations of the primary galaxy during the 
accretion (e.\ g.\ , triggering of spiral and bar distortions, redistribution 
of disk material), in addition to the deposition of accreted mass.\\

Low-density satellites that disrupt during their decay cause systematic 
structural transformations in the primary galaxy. It evolves towards higher 
$B/D$, higher $n$, higher $\sigma _0$, higher $h_D$ and lower 
$\mu _{0,Disk}$, all following trends similar to observations. A complete 
matching to present day galaxies may require the contribution of 
dissipative gas and star formation processes. \\

The infall of small, collapsed baryonic clumps is an inherent ingredient 
of galaxy 
formation models based on CDM \citep{White78}.  
The models presented 
here might indicate that such infall drives pre-existing disks to a 
secular evolution toward higher $B/D$ and $n$, through the inflow of primary 
disk material to the center via transitory bars, 
the deposition of 
satellite material rebuilding the exponential disk and the re-distribution 
of material vertically by heating.  Secular evolution is currenly 
being discussed as the outcome of bar instabilities in the disk due 
to gas cooling \citep{Kormendy04}.  The present models could indicate 
that secular evolution can be due to satellite accretion as well.  
The latter might have been important at earlier galaxy ages.  \\

\end{document}